\documentclass[conference]{IEEEtran}
\IEEEoverridecommandlockouts
\usepackage{cite}
\usepackage{amsmath,amssymb,amsfonts}
\usepackage{graphicx}
\usepackage{textcomp}
\usepackage{xcolor}

\usepackage[ruled,linesnumbered]{algorithm2e}
\usepackage{algpseudocode}

\makeatletter
\newcommand{\linebreakand}{%
  \end{@IEEEauthorhalign}
  \hfill\mbox{}\par
  \mbox{}\hfill\begin{@IEEEauthorhalign}
}
\makeatother

\def\BibTeX{{\rm B\kern-.05em{\sc i\kern-.025em b}\kern-.08em
    T\kern-.1667em\lower.7ex\hbox{E}\kern-.125emX}}
\begin{document}

\title{Heimdall++: Optimizing GPU Utilization and Pipeline Parallelism for Efficient Single-Pulse Detection\\
 \thanks{*Corresponding author}
}

\author{
\IEEEauthorblockN{Bingzheng Xia}
\IEEEauthorblockA{\textit{Hangzhou Institute for Advanced Study} \\
\textit{University of Chinese Academy of Sciences} \\
Hangzhou, China \\
xiabingzheng23@mails.ucas.ac.cn}
\and
\IEEEauthorblockN{Zujie Ren*, Kuang Mao}
\IEEEauthorblockA{\textit{Zhejiang Lab} \\
Hangzhou, China \\
renzju@zju.edu.cn \\
maok@zhejianglab.org}
\and
\IEEEauthorblockN{Xiaoqian Li, Wenda Li, Shuibing He}
\IEEEauthorblockA{\textit{Zhejiang University} \\
Hangzhou, China \\
lixiaoqian@zhejianglab.org\\
12421279@zju.edu.cn\\
heshuibing@zju.edu.cn}
}
\maketitle

\begin{abstract}
With the increasing time and frequency resolution of modern radio telescopes and the exponential growth in observational data volumes, real-time single-pulse detection has become a critical requirement for time-domain radio astronomy. Heimdall, as a representative GPU-accelerated single-pulse search tool, offers substantial performance advantages over CPU-based approaches. However, its sequential execution model and resource contention in intermediate processing stages limit GPU utilization, leading to suboptimal throughput and increased computational latency. To address these limitations, we present Heimdall++, an optimized successor to Heimdall that incorporates fine-grained GPU parallelization, enhanced memory management, and a multi-threaded framework to decouple CPU-bound and GPU-bound processing stages. This design mitigates the GPU stall problem and improves end-to-end efficiency. We evaluated Heimdall++ on a system equipped with NVIDIA RTX 3080 Ti GPUs using both a single large-scale observational file and multiple files. Experimental results demonstrate that Heimdall++ achieves up to 2.66$\times$ speedup in single-file processing and 2.05$\times$ speedup in multi-file batch processing, while maintaining full consistency with the original Heimdall's search results.
\end{abstract}


\section{Introduction}
\IEEEPARstart{P}{ulsars} are rapidly rotating, highly magnetized neutron stars. Pulsar search primarily relies on the Fast Fourier Transform (FFT) in the frequency domain and the Fast Folding Algorithm \cite{staelin1969fast} in the time domain. These methods are highly effective for detecting periodic pulsars. However, certain classes of pulsars—such as nulling pulsars, intermittent pulsars, and Rotating Radio Transients (RRATs)—exhibit irregular pulse sequences and are better detected through single-pulse searches. A fast radio burst (FRB) \cite{lorimer2007bright} is a millisecond-duration radio pulse with extremely high dispersion measures, likely originating from extragalactic sources, and can only be detected via single-pulse searches.

A major challenge in modern radio astronomy is managing the unprecedented data volumes generated by wide-band, multi-beam receivers, with data rates from current and upcoming facilities reaching tens to hundreds of terabits per second \cite{dewdney2015ska1, twidle2019impossible, schinckel2012australian, amiri2018chime}. Such massive data streams render long-term storage infeasible, necessitating real-time processing \cite{price2020real, williamson2024optimising}. This requirement demands not only large-scale computing hardware but also highly efficient single-pulse search software that is optimized to exploit available hardware resources fully.

As one of the most widely used Graphical Processing Units (GPU) accelerated single-pulse search tools, Heimdall\cite{barsdell2012accelerating} offers substantial performance advantages over CPU-based alternatives such as PRESTO  \cite{ransom2011presto} and SIGPROC \cite{lorimer2011sigproc}. Nevertheless, with the rapid growth of observational data, Heimdall’s computational bottlenecks have become increasingly prominent. In large dispersion measure (DM) and high-volume search scenarios, Heimdall often suffers from low GPU utilization, leading to prolonged processing times and inefficient hardware resource utilization.

Through systematic analysis, we identified several performance limitations in Heimdall. Firstly, the processing of dedispersed time series entails extensive loops across multiple stages, namely \textit{Baseline Removal}, \textit{Normalization}, \textit{Matched Filtering}, and \textit{Peak Detection}. These stages are predominantly executed in a sequential manner, thereby impeding the full utilization of GPU parallelism. Secondly, frequent data transfers between host and device over PCI Express (PCIe), especially for intermediate results of the dedispersion stage, cause GPU idleness, wasting valuable computational cycles. This phenomenon is referred to as the GPU stall problem \cite{agrawal2024taming}. Thirdly, inefficient memory access patterns, such as uncoalesced global memory accesses, further reduce throughput by underutilizing GPU cores. In multi-file scenarios, pipeline creation and execution are performed sequentially on the CPU and GPU, respectively, leading to GPU idle time during CPU processing. Although multi-process concurrency is often adopted to improve resource utilization, Heimdall’s implementation introduces significant resource contention, frequently causing process interruptions and limiting scalability in large-scale data processing.

To address these challenges, we propose Heimdall++, an end-to-end optimized redesign of the Heimdall pipeline that systematically enhances GPU utilization and throughput. At its core, Heimdall++ employs a fine-grained parallelization strategy: the DM trials loop, originally executed sequentially, is decomposed into independent tasks distributed across multiple CPU threads and CUDA streams. This enables concurrent kernel execution, increases streaming multiprocessor occupancy, and allows the degree of parallelism to be tuned to the target GPU’s capabilities. To minimize data movement overhead, Heimdall++ leverages CUDA Unified Memory for intermediate results, eliminating explicit host–device data copies while automatically managing memory residency across the CPU–GPU boundary. Furthermore, memory-bound stages such as candidate clustering are refactored to exploit GPU shared memory and coalesced access patterns, reducing global memory traffic by an order of magnitude. For multi-file workloads, Heimdall++ extends its efficiency through a multi-threaded, pipelined execution framework. Pipeline creation (a CPU-bound phase involving I/O and metadata setup) is decoupled from GPU-accelerated processing and connected via thread-safe task queues. This design enables overlapping of CPU and GPU activities across files, effectively masking CPU-side latency and mitigating the inherent GPU stall problem of sequential execution. A shared device-memory allocator further reduces contention during concurrent DM trial processing, ensuring scalable performance even under high thread counts. Collectively, these design choices enable Heimdall++ to sustain higher concurrency and improved scalability under different hardware constraints, particularly in large-scale survey workloads involving numerous files.

\section{RELATED WORK}

\subsection{Single-Pulse Search Tools}

Several software toolkits have been developed to facilitate single-pulse search in pulsar surveys. PRESTO and SIGPROC are widely used CPU-based toolchains designed for the initial analysis of high-time-resolution pulsar data. Another CPU-based single-pulse search tool is TransientX \cite{men2024transientx}. These packages offer foundational functionality for dedispersion, radio frequency interference (RFI) mitigation, and candidate generation but rely heavily on traditional serial or limited multi-threaded CPU computation, limiting their scalability with the exponential growth of observational data volumes.

Leveraging the superior parallel processing capabilities and memory bandwidth of GPU, several GPU-accelerated tools have been developed for pulsar and transient searches. The AstroAccelerate software package implements GPU-accelerated Fourier-domain acceleration search and harmonic sum-based periodicity search \cite{adamek2020single}. Heimdall is another prominent GPU-accelerated pipeline specifically developed for single-pulse detection. By exploiting the massive parallelism of modern GPUs, Heimdall achieves significantly higher processing speeds compared to its CPU-based counterparts, making it well-suited for real-time processing of large-scale survey data.

Previous studies have demonstrated the efficacy of GPU acceleration in pulsar and transient searches. For example, Sclocco et al. \cite{sclocco2016real} accelerated dedispersion on GPUs, reducing processing time by more than 50\%. You et al. \cite{you2021gpu} ported the dedispersion stage of PRESTO to GPUs, achieving speedups of 120$\times$ over the serial CPU version and 60$\times$ over the MPI-parallelized CPU version. More recently, Mao et al. \cite{mao2025prestozl} developed PrestoZL, a GPU-accelerated end-to-end pulsar search tool that introduces fine-grained GPU parallelization and a pipelined CPU–GPU execution framework to mitigate memory-bound bottlenecks and GPU stall issues. PrestoZL achieves up to a 56.38$\times$ speedup over the CPU-based PRESTO implementation while ensuring result consistency. These efforts underscore the critical role of GPU parallelism in addressing the burgeoning data volumes generated by radio astronomy. However, most existing GPU optimization work has focused on isolated stages (e.g., dedispersion) or periodic pulsar searches, leaving systematic GPU optimization of end-to-end single-pulse search pipelines relatively underexplored—a gap that our work aims to address.

\subsection{Pipeline Stages in Heimdall}

\begin{figure}[!t]
\centering
\includegraphics[width=0.8\linewidth]{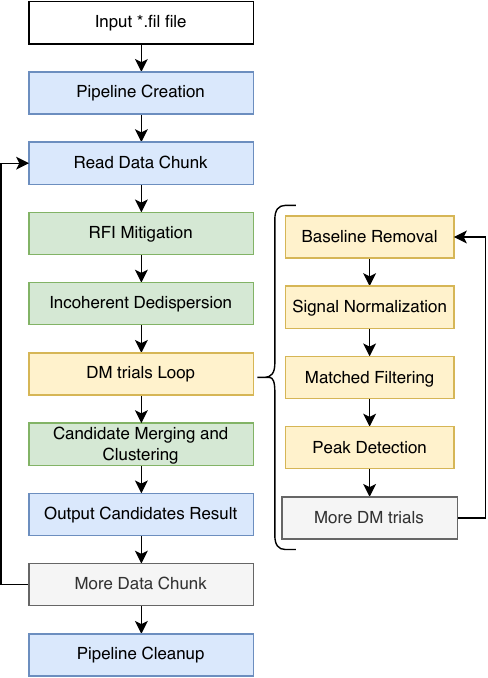}
\caption{Computational workflow of the original Heimdall pipeline, highlighting its sequential execution within the dispersion measure trial loop.}
\label{fig:heimdall_pipeline}
\end{figure}

As illustrated in Fig.~\ref{fig:heimdall_pipeline}, Heimdall implements a single-pulse search pipeline that follows the theoretical framework for transient detection \cite{cordes2003searches}. The workflow consists of the following stages:

\textit{RFI Mitigation.} Radio frequency interference is a major impediment to transient detection. Heimdall employs a multi-level strategy to suppress both broadband interference (spanning all frequency channels) and narrowband interference (confined to specific channels). By masking and filtering corrupted regions, the pipeline preserves astronomical signals for subsequent processing stages.  

\textit{Incoherent Dedispersion.} As radio waves propagate through the ionized interstellar medium, dispersion introduces frequency-dependent delays, quantified by the dispersion measure:
\begin{equation}
\mathrm{DM} = \int_{0}^{d} n_e(l) \, dl ,
\end{equation}
where \(n_e(l)\) denotes the electron density along the line of sight. Heimdall performs a search over hundreds to thousands of trial DMs using the GPU-accelerated \texttt{dedisp} library, aligning channels in time and summing them to maximize the signal-to-noise ratio (S/N). Dedispersion is one of the most computationally intensive stages and a primary target for GPU acceleration.  

\textit{Baseline Removal} and \textit{Normalization.} Instrumental and environmental effects can introduce low-frequency drifts and amplitude variations. Heimdall removes these trends via moving average subtraction and normalizes the dedispersed signal using root-mean-square (RMS) scaling. These steps stabilize the noise level, enabling consistent thresholding in subsequent detection stages. 

\textit{Matched Filtering} and \textit{Peak Detection.} Since the intrinsic pulse width is unknown in advance, Heimdall applies a bank of boxcar filters with varying widths to the time series. Convolving with these filters enhances pulse-like features across a wide range of durations. Peaks exceeding a predefined S/N threshold are extracted, with each candidate characterized by its time, trial DM, filter width, and peak S/N.  

\textit{Candidate Merging} and \textit{Clustering.} To reduce redundancy, Heimdall merges candidates that appear at similar times and DMs but arise from different trial parameters. A clustering algorithm groups such detections, and the strongest candidate is retained as the representative event. This step significantly reduces the candidate list, improving interpretability for downstream inspection or classification.  

\section{Heimdall Bottlenecks Analysis}
This section presents a systematic investigation of the existing Heimdall search pipeline. Through detailed performance profiling, we identify critical bottlenecks across distinct processing stages, which form the foundation for subsequent end-to-end optimization of the Heimdall workflow.

\subsection{GPU Utilization Analysis} \label{sec:GPU util analysis}

\begin{figure}
    \centering
    \includegraphics[width=1\linewidth]{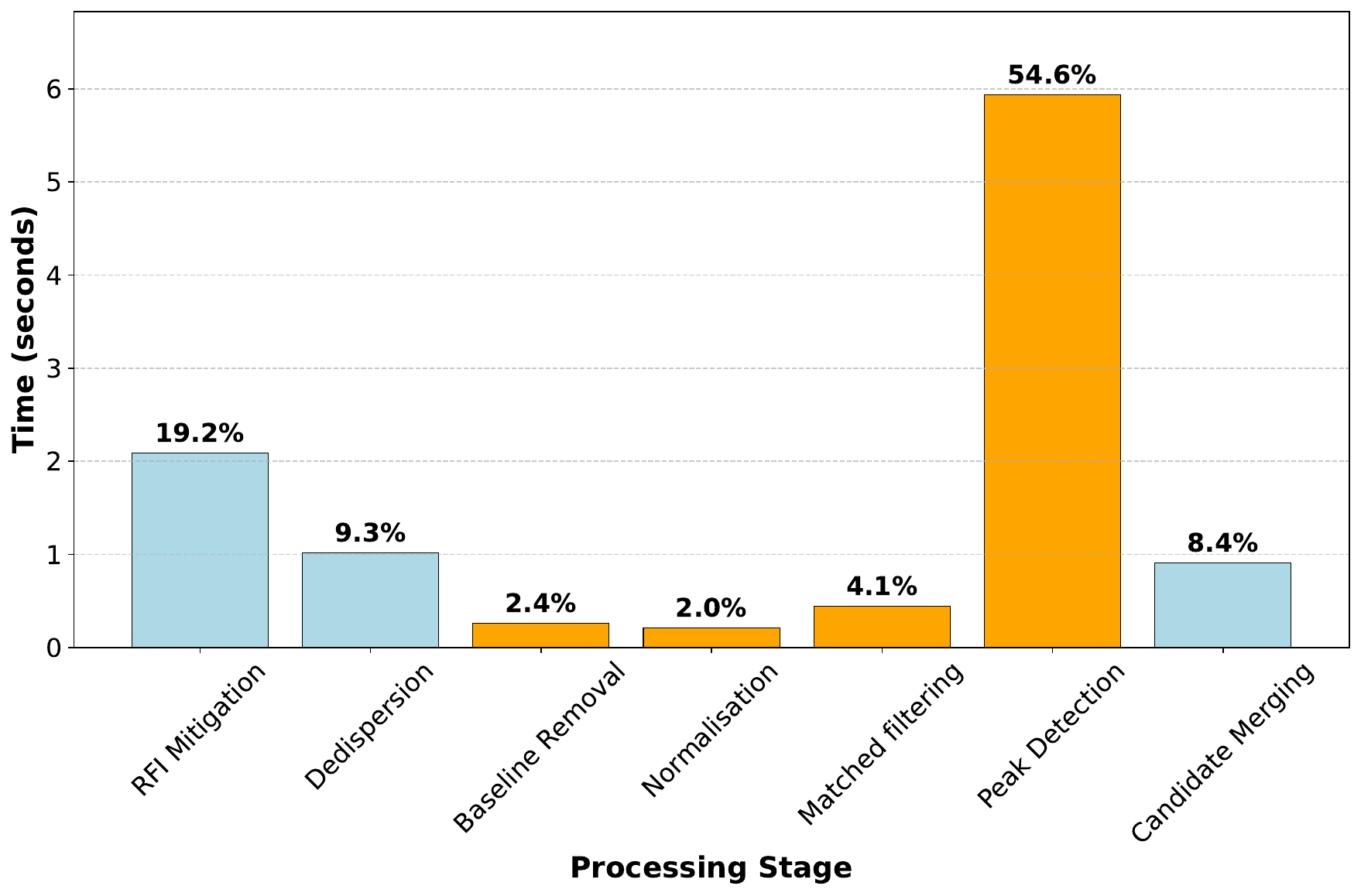}
    \caption{Stage-wise processing time in Heimdall. The yellow segment corresponds to the total runtime of the DM trials loop.}
    \label{fig:heimdall_time_bar_chart}
\end{figure}

\begin{figure}
    \centering
    \includegraphics[width=1\linewidth]{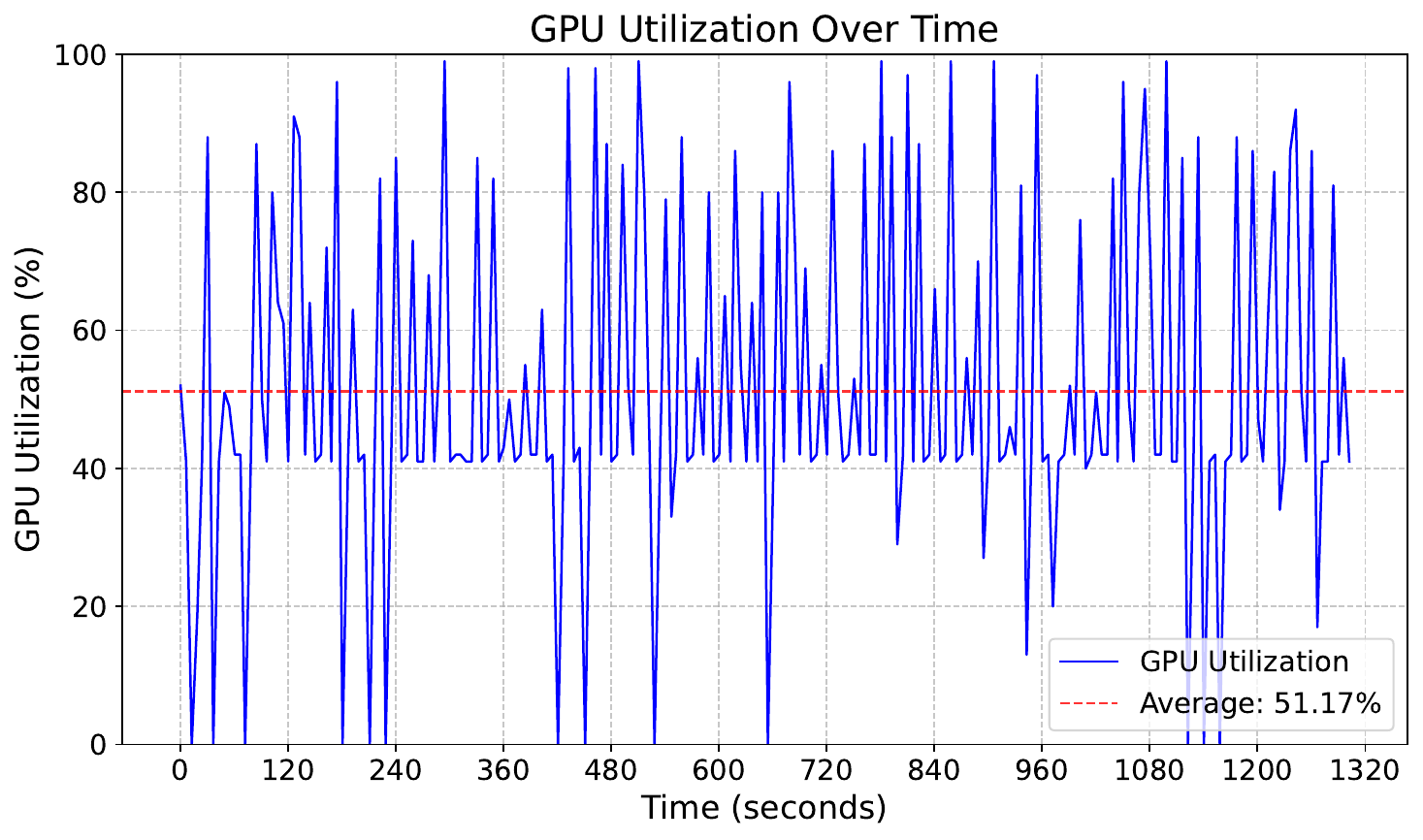}
    \caption{GPU utilization of Heimdall for processing a 1 GB input file.}
    \label{fig:heimdall_gpu_util}
\end{figure}

Heimdall exhibits low GPU utilization under typical single-pulse search workloads, indicating significant underutilization of available hardware resources. To quantify this inefficiency, we profiled a representative execution using a 1\,GB filterbank file processed as a single data chunk under default parameters; larger datasets repeat this per-chunk processing pattern without altering its behavior.

As shown in Fig.~\ref{fig:heimdall_gpu_util}, GPU utilization remains below 50\% for the majority of the execution, with an average of 51.2\%. Brief peaks occur only during pipeline initialization and finalization. This sustained underutilization directly increases end-to-end latency and represents a substantial waste of computational capacity.

The root cause lies in the serial execution of per-DM processing stages. By correlating Heimdall’s source code with NVIDIA Nsight Systems profiling data, we find that low-GPU-activity intervals align precisely with the compute-intensive stages highlighted in orange in Fig.~\ref{fig:heimdall_time_bar_chart}. Specifically, \textit{Baseline Removal}, \textit{Normalization}, \textit{Matched Filtering}, and \textit{Peak Detection} are executed sequentially across hundreds to thousands of DM trials generated during \textit{Incoherent Dedispersion}. Since each trial operates on a relatively small dedispersed time series, the per-iteration workload is insufficient to saturate the GPU massive parallelism. Consequently, the device remains underutilized throughout the DM trial loop.

To mitigate this bottleneck, concurrency across DM trials must be increased to improve GPU occupancy and reduce the total runtime of the looped processing stages.

\subsection{Data Transfer Analysis Between Host and Device} \label{sec:data transfer}

Excessive host--device data transfers over the PCIe bus constitute a major performance bottleneck in Heimdall, severely limiting GPU utilization. When processing a 1\,GB input file, Heimdall performs over 129,000 host-to-device and 220,000 device-to-host memory copy operations, resulting in a total data movement of approximately 7.85\,GB—nearly eight times the input size. This indicates repeated transfer of intermediate results across the PCIe interface during pipeline execution.

Since PCIe bandwidth is substantially lower than on-device memory bandwidth, these frequent transfers introduce significant latency. The GPU frequently stalls while waiting for data, leading to underutilization, degraded memory efficiency, and reduced end-to-end throughput. Code-level inspection combined with the GPU utilization profile in Fig.~\ref{fig:heimdall_gpu_util} shows that the majority of redundant transfers occur in and around the \textit{Dedispersion} stage. In the baseline implementation, both inputs and outputs of dedispersion reside in host memory. Consequently, data produced by upstream GPU stages are first copied back to the host for dedispersion and then retransferred to the device for downstream processing, resulting in costly PCIe round-trips.

This design is motivated by the need to handle a large number of DM trials, as the combined output from all these trials after dedispersion can exceed the GPU's available memory capacity.  Reducing the per-chunk data size is not a viable solution: smaller chunks underutilize GPU parallelism and increase redundant computation due to overlapping time segments required to preserve signal continuity across chunk boundaries. 

Therefore, our optimization goal is to minimize unnecessary data movement while ensuring that dedispersion outputs remain within device memory limits. By retaining sufficiently large chunk sizes, Heimdall++ sustains high GPU occupancy and avoids excessive recomputation, thereby improving end-to-end throughput.

\subsection{Stage-level Bottlenecks Analysis}

Building on the preceding system-level analysis, this subsection provides a fine-grained examination of individual processing stages in Heimdall to identify stage-specific performance bottlenecks and establish the foundation for end-to-end optimization.

The \textit{RFI Mitigation} stage suffers from redundant host-device memory transfers that degrade its efficiency. This stage includes optional broadband and narrowband components targeting spectrally broad and terrestrial interference, respectively. Broadband RFI, being undispersed in the interstellar medium, is typically detected using a single $\mathrm{DM}=0$ trial to suppress dispersed astrophysical signals and improve interference excision reliability. Despite requiring only one DM trial, the current implementation unnecessarily transfers both input and output data between host and device memory. Retaining these data entirely in device memory would eliminate PCIe traffic and reduce latency without compromising functionality.

The \textit{Peak Detection} stage dominates the pipeline runtime, accounting for over 50\% of Heimdall’s total execution time, as shown in Fig.~\ref{fig:heimdall_time_bar_chart}. This stage relies on NVIDIA’s Thrust library to execute primitives such as \texttt{reduce} and \texttt{inclusive\_scan}. Although each temporary buffer allocation incurs modest overhead, repeated invocation across hundreds to thousands of DM trials accumulates significant latency. A device-side memory pool that enables buffer reuse across trials would eliminate redundant \texttt{cudaMalloc}/\texttt{cudaFree} calls and amortize allocation costs over the entire pipeline. Extending this memory-reuse strategy to other memory-bound stages in the Thrust implementation would further enhance throughput.

The \textit{Candidate Merging} and \textit{Clustering} stage consolidates redundant detections by grouping nearby candidates in the three-dimensional parameter space of time, DM, and filter width, retaining the highest S/N candidate per cluster. However, the current implementation exhibits $O(N^2)$ complexity, as each candidate is compared against all others to establish cluster membership. This results in extensive non-coalesced global-memory accesses, rendering the stage memory-bound and increasingly latency-dominated as the candidate count $N$ grows. Refactoring the algorithm to better exploit the GPU memory hierarchy—by promoting coalesced access patterns and reducing global memory transactions—would alleviate bandwidth pressure and significantly accelerate this stage.

\subsection{Bottlenecks in Parallel Processing of Multiple Files}

In large-scale multi-file processing scenarios, it is common to process multiple filterbank files concurrently by launching parallel Heimdall processes on a single GPU. However, this approach incurs severe CUDA context contention: only one context can execute at a time, resulting in time-sliced kernel scheduling and frequent context switches. As a consequence, effective concurrency collapses, and overall GPU throughput degrades despite the presence of multiple active processes. This issue is particularly pronounced for small-file workloads. In these scenarios, the fine-grained nature of kernels causes context-switching and launch overheads to approach the actual computation time, leading to persistently low GPU utilization with significant fluctuations that prevent full occupancy of the device.

An additional bottleneck arises from pipeline serialization within each process. File I/O, CPU preprocessing, and GPU computation are executed strictly sequentially without stage overlap. This absence of I/O–compute concurrency results in intermittent idling of both CPU and GPU resources, further constraining end-to-end throughput even under multi-process execution. Collectively, these limitations render the current design latency-bound and poorly scalable, necessitating a redesigned architecture to eliminate context contention and enable deep overlap between data movement and computation.

\section{Optimization Design Principles in Heimdall++}

In this section, building on the preceding analysis of Heimdall’s bottlenecks, we present a redesigned processing pipeline, hereafter referred to as Heimdall++. Heimdall++ delivers more efficient GPU execution and accelerates the end-to-end workflow, achieving higher throughput on the same hardware while preserving result equivalence with the original Heimdall. 

\begin{figure}
    \centering
    \includegraphics[width=0.9\linewidth]{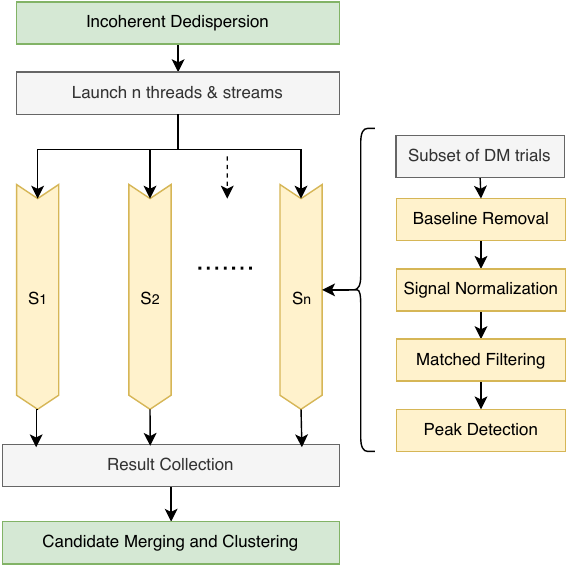}
    \caption{Heimdall++ multi-stream parallelization architecture. Here, \( S_1, S_2, \dots, S_n \) denote individual CUDA streams, each assigned a subset of DM trials. The DM trial loop is distributed across multiple CPU threads and CUDA streams to enable concurrent execution of \textit{Baseline Removal}, \textit{Normalization}, \textit{Matched Filtering}, and \textit{Peak Detection}.}
    \label{fig:heimdall++_pipeline}
\end{figure}

\subsection{Increasing Parallelism}

Heimdall processes DM trials in strict serial order, which severely limits GPU utilization and increases end-to-end latency. To overcome this bottleneck, we introduce a fine-grained parallelization strategy that exploits CUDA multi-stream execution and a multi-threaded shared device-memory allocator to enable concurrent processing of DM trials and efficient buffer reuse.

\begin{algorithm}[t]
\caption{Single-Pulse Search Procedure in Heimdall++}
\label{alg:heimdall++}
\SetKwProg{Fn}{Function}{}{end} 
\textbf{Load Data and Preprocess (in CPU)}\;
\textbf{RFI Mitigation (in GPU)}\;
\quad Apply broadband and narrowband \textit{RFI mitigation}\;
\textbf{Dedispersion (in GPU)}\;
\quad Perform GPU-based \textit{Incoherent Dedispersion} over trial DMs\;

\textbf{Parallel Candidate Search (both CPU and GPU)}\;
\For{each OpenMP thread $t \in [0, T-1]$}{
    Assign subset of DM trials: $DM_t = \{DM_i \mid i \equiv t \pmod{T}\}$\;
    Create CUDA stream $S_t$\;
    \For{each $DM \in DM_t$}{
        Allocate memory via shared device-memory allocator\;
        Launch kernels asynchronously in $S_t$: \\
        \quad \textit{Baseline Removal}, \textit{Normalization}, \textit{Matched Filtering}, \textit{Peak Detection}\;
        Collect candidates into thread-local buffer\;
    }
}
Synchronize all threads and aggregate candidates from all buffers\;
\textbf{Candidate Merging and Clustering (GPU)}\;
\quad Group candidates in the time–DM–filter width space\;
\quad Retain the candidate with maximum S/N per cluster\;
\end{algorithm}

As illustrated in Fig.~\ref{fig:heimdall++_pipeline} and Algorithm~\ref{alg:heimdall++}, after the \textit{Dedispersion} stage completes, an OpenMP thread team is launched with a user-specified number of threads $T$. Each thread partitions the DM trial space using its thread ID and computes the memory indices required to access its assigned dedispersed time series. On the GPU side, every host thread creates a dedicated CUDA stream, and all subsequent kernels—\textit{Baseline Removal}, \textit{Normalization}, \textit{Matched Filtering}, and \textit{Peak Detection}—are launched asynchronously within that stream. This decoupling enables asynchronous execution between CPU orchestration and GPU computation, reducing synchronization overhead.

The use of multiple CUDA streams facilitates true parallel execution of independent DM trials, which are otherwise processed sequentially in the original Heimdall pipeline. By launching each trial as an independent kernel in its own stream, we increase streaming multiprocessor occupancy and reduce idle cycles. Moreover, overlapping kernel execution across streams mitigates per-kernel launch overhead and improves overall throughput through better hardware utilization.

Concurrent execution of Thrust-based primitives across streams increases the demand for temporary device memory. To avoid the performance penalty of repeated \texttt{cudaMalloc}/\texttt{cudaFree} calls, we replace Thrust’s default allocator with a custom multi-threaded shared device-memory allocator. This allocator maintains two global queues during program execution: an allocated-block queue and a free-block queue, which track in-use and available memory blocks. When a thread requests memory, the allocator first aligns the requested size and searches the free-block queue. If no suitable block is found, it allocates a new block from device memory. Upon release, blocks are not deallocated immediately, but returned to the free queue for reuse by other threads. Thread-safe access to these queues is ensured via C++ reader-writer lock mechanisms, which prevent race conditions while supporting high-throughput concurrent allocation and deallocation.

\subsection{Reduce Host--Device Data Transfers}

Heimdall++ eliminates redundant host--device data transfers by leveraging CUDA Unified Memory, thereby reducing PCIe traffic while avoiding GPU memory exhaustion.

In the original Heimdall implementation, both inputs and outputs of the \textit{Dedispersion} stage reside in host memory. An initial optimization that retained all dedispersed data in device memory proved infeasible: the output size scales linearly with the user-specified DM range, and large DM intervals or chunk sizes can easily exceed GPU memory capacity. Furthermore, downstream stages require additional buffers for intermediate results, exacerbating memory pressure. Storing all data on-device thus risks out-of-memory failures and program termination.

To address this trade-off between data locality and memory capacity, Heimdall++ adopts CUDA Unified Memory, which provides a unified virtual address space accessible from both CPU and GPU. The CUDA runtime system and hardware automatically migrate memory pages on demand between host and device, enabling the working set to exceed the physical GPU memory limit. We refactored the \textit{Dedispersion} stage to read input data directly from device memory and write dedispersed outputs into Unified Memory. When the total output footprint exceeds available GPU memory, less-frequently accessed pages are transparently migrated to host memory.

Unlike the original design, which performs bulk, explicit \texttt{cudaMemcpy} operations for every intermediate result, this approach reduces data movement through fine-grained, on-demand memory migration. As a result, PCIe bandwidth is used more efficiently, minimizing GPU idle time caused by data transfer stalls. Experimental results show that the overhead of Unified Memory management is negligible relative to the achieved performance improvements. Consequently, end-to-end throughput is significantly improved without sacrificing robustness under large DM ranges or high-resolution observational data.

\subsection{Multi-File Pipeline Parallelism Design}
Heimdall++ addresses the inefficiencies of multi-file processing by introducing a two-stage pipelined architecture that decouples CPU-bound setup from GPU-bound computation, thereby mitigating the GPU stall problem and eliminating CUDA context contention inherent in multi-process execution.

\begin{figure}
\centering
\includegraphics[width=1\linewidth]{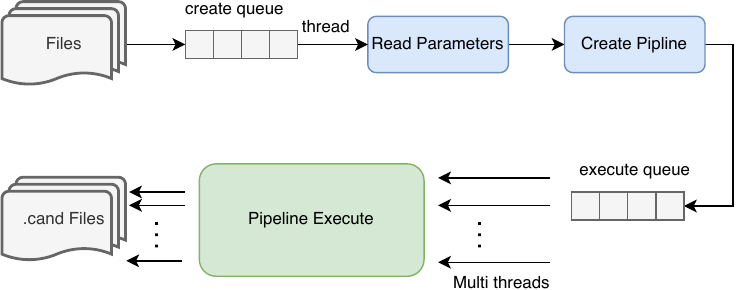}
\caption{Pipeline parallel framework of Heimdall++}
\label{fig:heimdall_PL}
\end{figure}

Fig.~\ref{fig:heimdall_PL} illustrates the pipeline-parallel framework of Heimdall++. In large-scale survey pipelines, processing numerous small filterbank files sequentially leads to significant GPU idle time, as each file undergoes independent pipeline initialization, I/O, and execution. To overcome this, Heimdall++ restructures the workflow into two asynchronous concurrent stages connected by thread-safe task queues.

The first stage, \textbf{Pipeline Creation}, is executed on the CPU and handles all file-level initialization tasks. For each input filterbank file, it reads metadata (e.g., center frequency, bandwidth, sampling interval, and beam index), configures Heimdall search parameters, allocates Unified Memory buffers, and constructs a lightweight \texttt{PipelineTask} object that encapsulates the file context and execution plan. Upon completion, the task is enqueued into a \textit{creation queue}, enabling immediate handoff to the next stage without blocking.

The second stage, \textbf{GPU Execution}, consumes tasks from an \textit{execution queue} and performs the core signal processing pipeline: data chunks are streamed into GPU memory, \textit{Incoherent Dedispersion} is applied across DM trials, and candidate events are identified via \textit{Matched Filtering} and \textit{Peak Detection}. Critically, this stage overlaps I/O and computation through double buffering and asynchronous memory transfers, ensuring the GPU remains continuously occupied. By executing all files within a single process using a multi-threaded runtime, Heimdall++ avoids the overhead of CUDA context switching and inter-process resource contention that plagues traditional multi-process approaches.

Lock-free or mutex-protected task queues mediate inter-stage communication, guaranteeing thread safety while minimizing synchronization overhead. This decoupling enables pipeline creation for the next file to proceed while the GPU is still processing the current one, effectively masking CPU-side latency and sustaining high device utilization across batch workloads. As a result, the system achieves higher throughput, better scalability, and more predictable performance under realistic multi-file observational scenarios.

\section{EXPERIMENTS AND RESULTS}
In this section, we present a performance evaluation of Heimdall++, using the original Heimdall as the baseline for comparison. We first examine the acceleration achieved by Heimdall++ across different pipeline stages, with particular emphasis on GPU utilization analysis. Subsequently, we conducted experiments under two representative scenarios. (i) large-file processing, corresponding to long-duration continuous observational data, and (ii) batch processing of multiple files. Finally, we validate that the search results produced by Heimdall++ are consistent with those of the original Heimdall, ensuring that the optimization improves computational efficiency without compromising scientific accuracy.

\subsection{Experimental Setting}
\textbf{Experiment Environment.} All experiments were conducted on a workstation equipped with an NVIDIA GeForce RTX 3080 Ti GPU and a 12th Gen Intel(R) Core(TM) i9-12900K CPU, running Ubuntu 20.04.6 LTS. GPU programming was carried out using the NVIDIA CUDA Toolkit 11.8.  

\textbf{Experimental Data.} For the stage-level performance analysis, we used the same dataset as in our previous Heimdall profiling. The specific PSRFITS file employed is \texttt{J0528\_2200\_arcdrift-M01\_0009.fits}, with a size of 1 GB, obtained from the Commensal Radio Astronomy FAST Survey (CRAFTS) \cite{li2018fast}. This file was converted into the filterbank format using the \texttt{psrfits2fil.py} tool in PRESTO to ensure compatibility with Heimdall.  

To evaluate performance on long-duration observations, we further used the archived globular cluster M5 (NGC 5904) \cite{freire2024gcpsr} dataset from FAST, which consists of 282 combined FITS files representing a 30-minute observation. After conversion using \texttt{psrfits2fil.py}, we obtained a 142 GB filterbank file. In this experiment, the dispersion measure range was set from 0 to 1000~cm$^{-3}$, and the chunk size for each processing iteration was fixed at 256K samples.  

For multi-file batch processing experiments, we used the FRB20201124 subset of the FAST-FREX dataset \cite{Guo_2025}. This dataset is based on FAST observations and contains 125 FITS sample files, each of size 488 MB, with one FRB signal per file. These files were converted into filterbank format while preserving file size. 

This combination of single-file and multi-file workloads provides a comprehensive evaluation of Heimdall++ under realistic observational conditions. 

\subsection{Stage-level Performance Comparison}

To evaluate the acceleration achieved by Heimdall++ at each processing stage, we used the same dataset as in our previous performance analysis of Heimdall. The specific PSRFITS file employed for testing is J0528\_2200\_arcdrift-M01\_0009.fits, with a size of 1 GB, obtained from the CRAFTS sky coverage project \cite{li2018fast}. We converted this file into the filterbank format using the Psrfits2fil.py tool in PRESTO to make it compatible with Heimdall processing.

\begin{figure}[!t]
\centering
\includegraphics[width=1\linewidth]{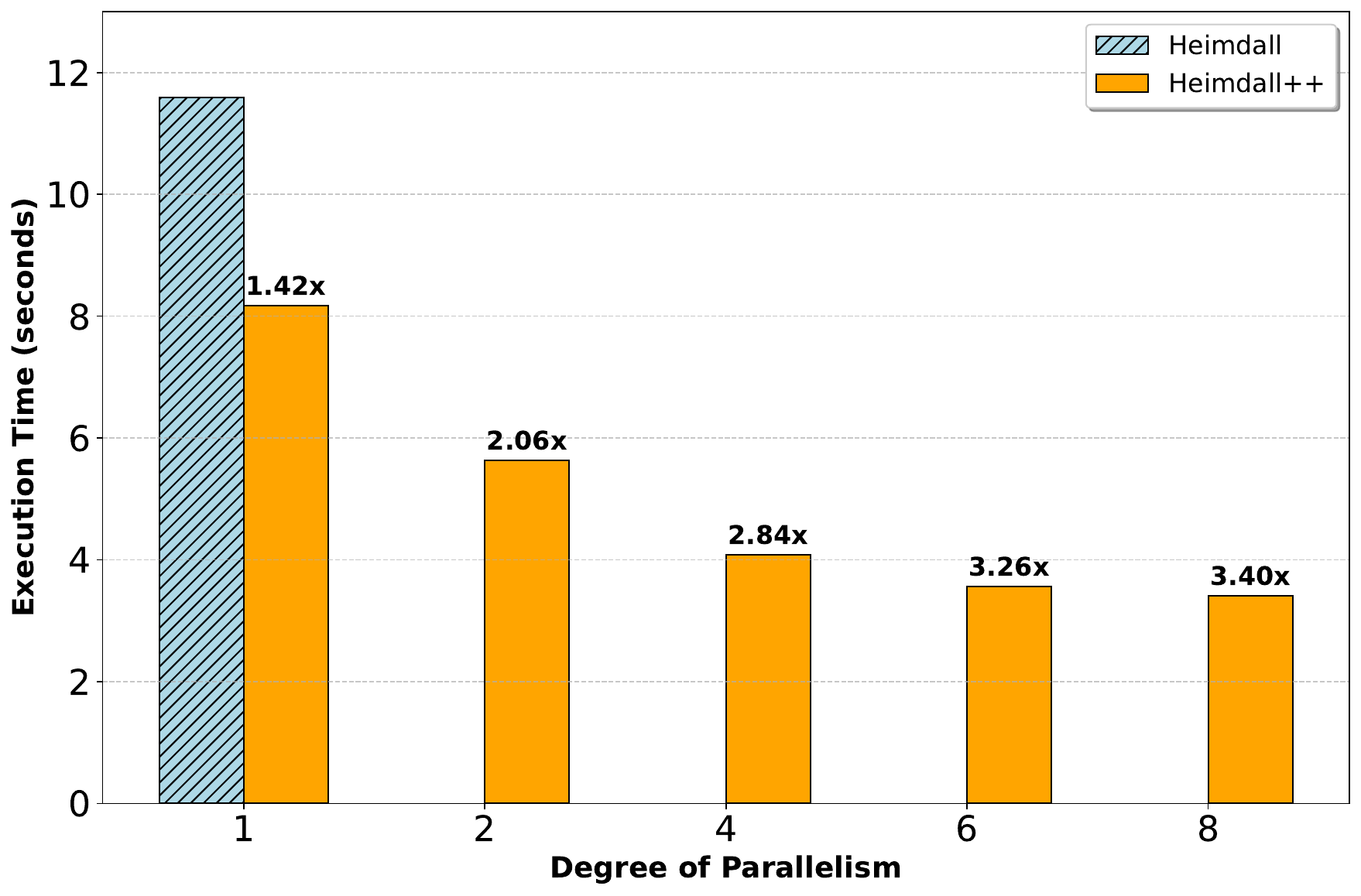}
\caption{Performance comparison between Heimdall++ and Heimdall under different parallelisms}
\label{fig:heimdall cmp heimdall++}
\end{figure}

\begin{figure}[!t]
\centering
\includegraphics[width=1\linewidth]{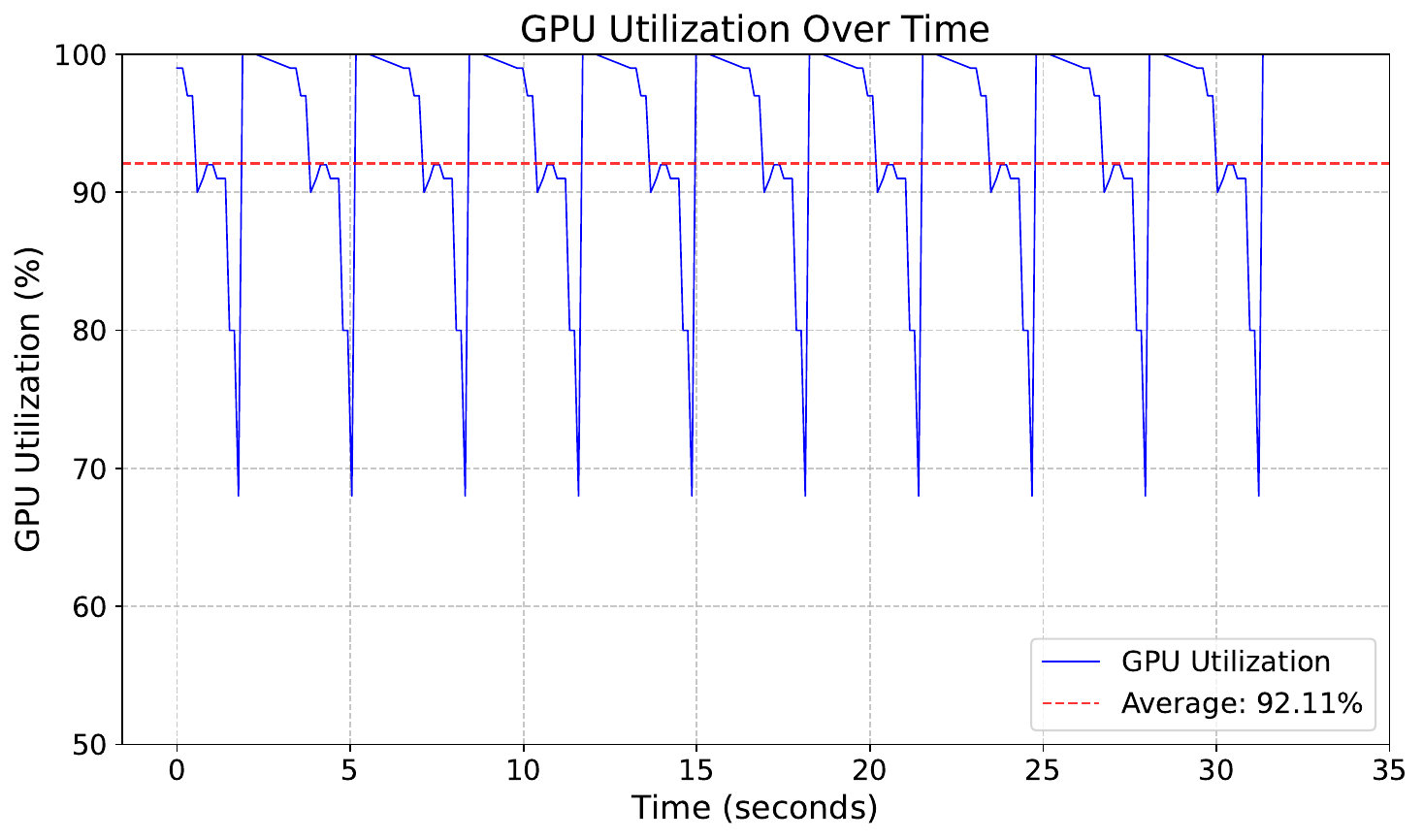}
\caption{GPU utilization of Heimdall++ when processing one data chunk with the parallelism is 8}
\label{fig:Heimdall++_gpu_t8}
\end{figure}

\begin{table}[!t]
\renewcommand{\arraystretch}{1.3}
\caption{Total volume of host--device data transfer when processing a 1GB-sized observational file}
\label{tab:cuda-mem-comparison}
\centering
\begin{tabular}{lcc}
\hline
\hline
Transfer direction & Heimdall & Heimdall++ \\
\hline
Host-to-Device & 5.19\,GB & 716.87\,MB \\
Device-to-Host & 2.66\,GB & 480.75\,MB \\
\hline
Total volume & 7.85\,GB & 1.17\,GB \\
\hline
\hline
\end{tabular}
\end{table}

As shown in Fig.~\ref{fig:heimdall cmp heimdall++}, we compared the pipeline execution time of Heimdall and Heimdall++ under our experimental setup. The results demonstrate that as the degree of parallelism increases from 1 to 8, Heimdall++'s processing speed relative to Heimdall steadily improves, with the speedup increasing from 1.42$\times$ to 3.40$\times$. This confirms the effectiveness of our design. When parallelism is set to 1, the execution flow of Heimdall++ is essentially identical to that of the original Heimdall, both performing serial processing with a single thread and a single stream. Nevertheless, owing to the optimizations introduced in other stages of Heimdall++, even under this configuration, the system achieves more than a 40\% performance improvement over Heimdall. As the parallelism increases to 2 and 4, the speedup grows significantly, reaching 2.06$\times$ and 2.84$\times$, respectively. Further increasing the parallelism to 6 and 8 yields additional gains, but the rate of improvement diminishes. This indicates that at higher levels of parallelism, the loop computation approaches the computational limits of the GPU hardware, such that device capability rather than algorithm design becomes the dominant constraint on performance scalability.

To more clearly illustrate the GPU utilization during the execution of Heimdall++, we present in Fig.~\ref{fig:Heimdall++_gpu_t8} the utilization profiles obtained when processing with parallelism parameters of 8, respectively. As shown in Fig.~\ref{fig:Heimdall++_gpu_t8}, with parallelism set to 8, the average GPU utilization of Heimdall++ is approximately 92\%, higher than the 47\% observed for Heimdall in Fig.~\ref{fig:heimdall_gpu_util}. The comparison further indicates that Heimdall++ eliminates the periods of GPU stalls present in Heimdall, primarily because the redesigned pipeline reduces idle time by avoiding excessive data transfers between host and device memory over PCIe. Nonetheless, the extended low-utilization phase during the DM trials loop still prolongs the overall runtime. This improvement demonstrates that distributing DM trials loop iterations across multiple threads enables more effective exploitation of GPU parallelism. The brief utilization drop near the end of the loop reflects uneven completion times among threads, where some threads finish earlier than others, temporarily reducing active workload on the device. Overall, these results confirm the effectiveness of increasing parallelism to enhance GPU resource utilization in Heimdall++.

\begin{figure}[!t]
\centering
\includegraphics[width=1\linewidth]{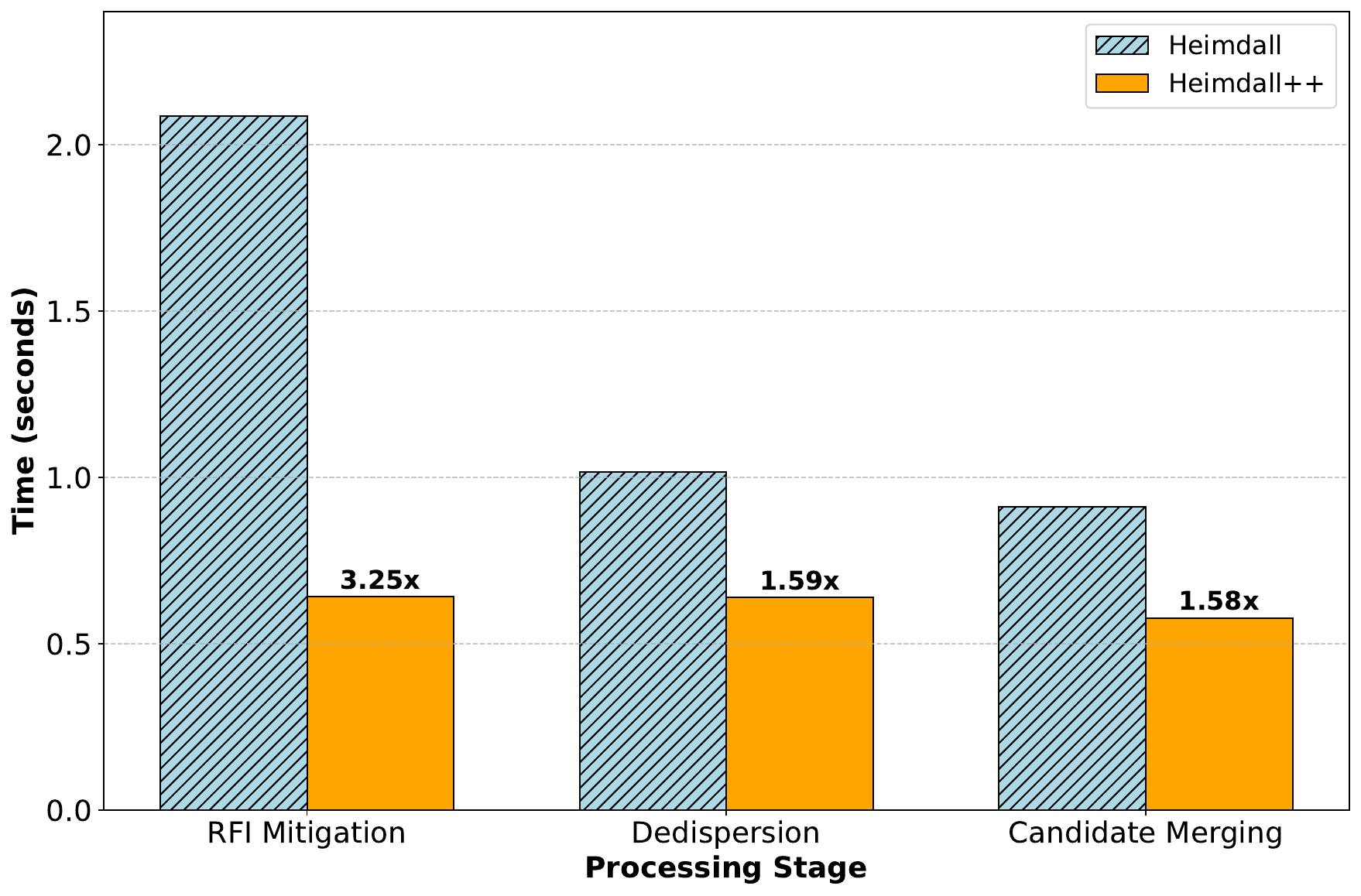}
\caption{Stage-wise performance comparison between Heimdall and Heimdall++ under single-thread mode}
\label{fig:stage-wise Heimdall++ single-thread}
\end{figure}

\begin{figure*}[!t]
\centering
\includegraphics[width=1\linewidth]{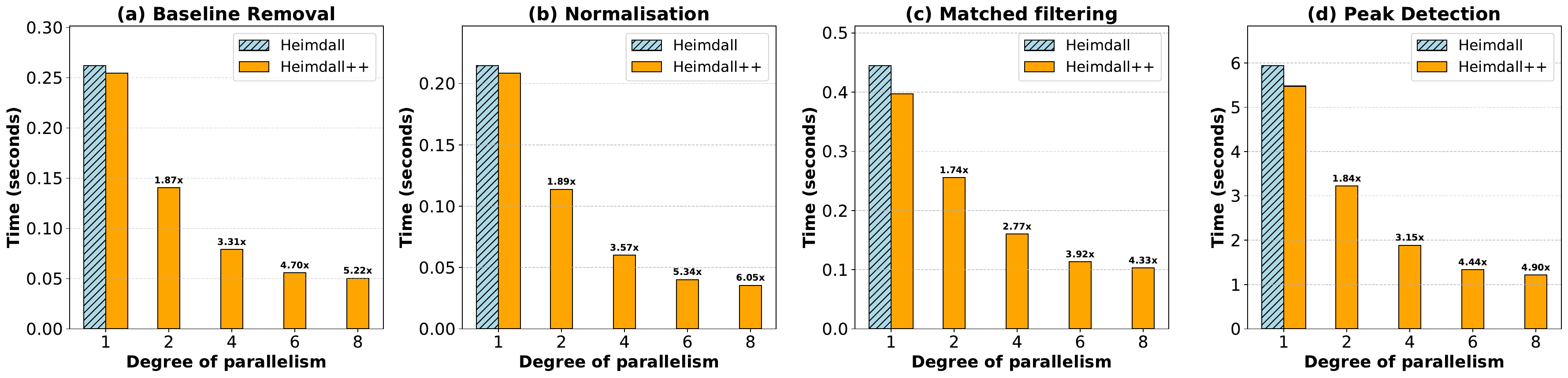}
\caption{Stage-wise performance comparison between Heimdall and Heimdall++ across different degrees of parallelism}
\label{fig:stage-wise Heimdall++ multi-thread}
\end{figure*}

To complement the overall comparison presented in the previous section, we further conducted a stage-wise performance analysis of Heimdall and Heimdall++. Fig.~\ref{fig:stage-wise Heimdall++ single-thread} presents the execution time comparison of the \textit{RFI Mitigation}, \textit{Dedispersion}, and \textit{Candidate Merging} stages between Heimdall and Heimdall++. As shown in panels (a) and (b), the performance of \textit{RFI mitigation} and Dedispersion improved by factors of 3.25$\times$ and 1.59$\times$, respectively. These gains primarily result from our optimization of redundant data transfers between host and device memory. By leveraging CUDA global memory, intermediate results between stages are preserved on the device, thereby avoiding explicit host–device copies and improving overall efficiency. As quantified in Table \ref{tab:cuda-mem-comparison}, the total data movement for a 1\, GB input file is reduced from 7.85\, GB in Heimdall to just 1.17\, GB in Heimdall++, a 6.7 times decrease. This dramatic reduction minimizes GPU idle time caused by data stalls and enables more efficient utilization of computational resources. As shown in panel (c), we redesigned the \textit{Clustering} algorithm used in the \textit{Candidate Merging} stage of Heimdall++. By exploiting GPU shared memory to enable coalesced access to global memory and simultaneously reducing the number of global memory transactions, we alleviated the memory-bound nature of the \textit{Clustering} procedure and thereby improved execution efficiency. Moreover, the benefits of this optimization become increasingly significant as the number of candidates grows, highlighting its scalability for large-scale survey workloads.

\begin{table}[!t]
\renewcommand{\arraystretch}{1.3}
\caption{Comparison of cudaMalloc call counts between Heimdall and Heimdall++}
\label{tab:malloc}
\centering
\begin{tabular}{lccc}
\hline
\hline
Implementation & cudaMalloc Calls Counts & Reduction Ratio \\
\hline
Heimdall    & 198,392 & – \\
Heimdall++  & 4,810   & 41.2$\times$ fewer \\
\hline
\hline
\end{tabular}
\end{table}

Fig.~\ref{fig:stage-wise Heimdall++ multi-thread} illustrates the speedup of Heimdall++ over the original Heimdall in the DM trial loop stages, including \textit{Baseline Removal}, \textit{Normalization}, \textit{Matched Filtering}, and \textit{Peak Detection}, under different degrees of parallelism. The results show that increasing the level of parallelism gradually improves the acceleration ratio of Heimdall++, thereby enhancing computational efficiency within the loop. At parallelism 8, Heimdall++ achieves a maximum speedup of 6.05$\times$ in \textit{Normalization} (Fig. 9(b)) and a minimum of 4.33$\times$ in \textit{Matched Filtering} (Fig. 9(c)). This improvement is further supported by our multi-threaded shared device-memory allocator mechanism. As shown in Table \ref{tab:malloc}, this custom memory reuse scheme reduces the number of \texttt{cudaMalloc} calls by a factor of 41.2 compared to Heimdall, ensuring efficient multi-thread parallelism and avoiding performance degradation caused by frequent memory allocation contention.

These stage-level evaluations confirm that the end-to-end acceleration of Heimdall++ originates from systematic improvements across the pipeline. By reducing redundant memory operations, introducing efficient buffer reuse, and enabling fine-grained parallel execution, Heimdall++ achieves higher throughput and better GPU utilization than the original Heimdall across both single-threaded and multi-threaded regimes.

\subsection{Large File Processing Performance}
\begin{figure}[!t]
\centering
\includegraphics[width=1\linewidth]{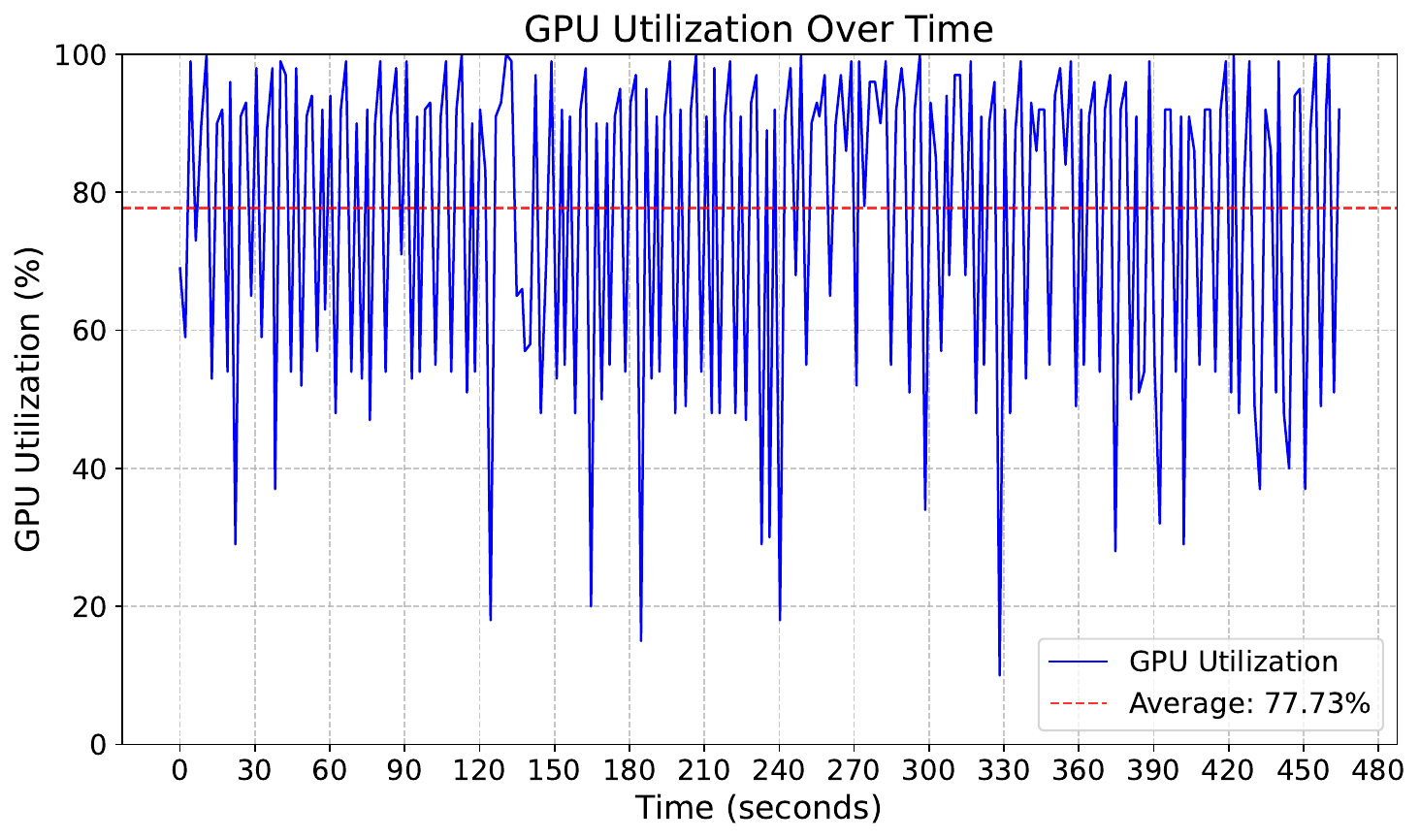}
\caption{Heimdall++ GPU utilization over time in processing M5}
\label{fig:heimdallpp_largefile_util}
\end{figure}

\begin{figure}[!t]
\centering
\includegraphics[width=1\linewidth]{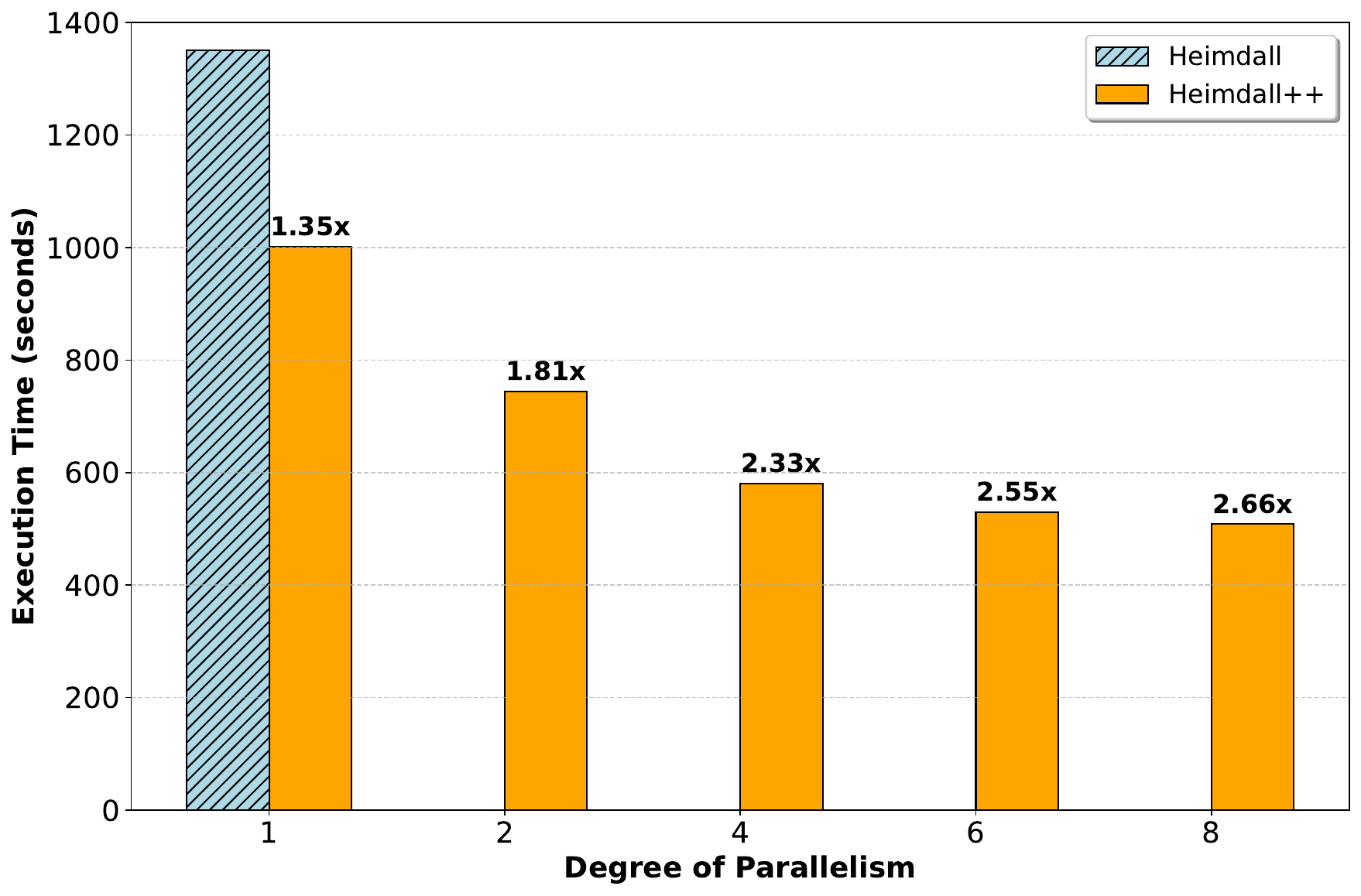}
\caption{Performance comparison between Heimdall and Heimdall++ in various degrees of parallelism for processing M5}
\label{fig:heimdallpp_largefile_speedup}
\end{figure}

To further evaluate the scalability of Heimdall++ in realistic large-scale observational scenarios, we conducted experiments on a 142 GB filterbank file, representing long-duration radio survey data. The original Heimdall processes such large files sequentially, where each data chunk is loaded from disk, transferred to GPU memory, and processed in isolation. This pipeline design inherently causes serial blocking between file I/O, host-device memory transfers, and GPU computation, leading to substantial idle time on the GPU and thus inefficient resource utilization.

We conducted the experiments using the default parameters of Heimdall, with the DM range set from 0 to 1000 cm$^{-3}$. For each run, the chunk consists of 256K samples. The performance of Heimdall and Heimdall++ was then compared under a single-process configuration with different degrees of parallelism.

As shown in Fig.~\ref{fig:heimdallpp_largefile_speedup}, the speedup of Heimdall++ relative to Heimdall increases steadily with higher degrees of parallelism, rising from 1.35$\times$ at parallelism 1 to 2.66$\times$ at parallelism 8. A comparison with Fig.~\ref{fig:heimdall cmp heimdall++} reveals that, under the same parallelism settings, the speedup achieved by Heimdall++ when processing the 142 GB M5 file is lower than that obtained for the 1 GB file. This difference arises because, in the 1 GB experiments, we excluded the time overhead of pipeline creation and data preprocessing in order to isolate the gains from computational optimizations within the pipeline. These stages incur a fixed cost that does not scale with parallelism, and thus their inclusion reduces the overall acceleration ratio in the large-file scenario. This more accurately reflects the speedup achievable in practical applications.

Moreover, for large-file processing, we employed a double-buffering strategy to asynchronously read sample data blocks, thereby overlapping I/O with computation across successive batches. This approach effectively mitigates GPU stalls and further improves end-to-end efficiency. The large-file experiments therefore provide strong evidence that Heimdall++ delivers substantial performance improvements over the original Heimdall, achieving multi-fold acceleration on the same hardware platform.

\subsection{Multi-file Processing Performance Comparison}
In practical applications, batch processing of astronomical observation files is a common scenario. A straightforward approach for handling multiple files is to employ multi-process concurrency. However, our experiments with Heimdall revealed that multi-process execution results in poor processing and parallel efficiency. Specifically, CPU-side overhead often leaves the GPU underutilized, while simply increasing the number of concurrent processes leads to frequent resource contention, reduced efficiency, and in some cases, process crashes.

By leveraging multi-threading to increase concurrency while avoiding excessive resource contention, Heimdall++ enables more efficient utilization of hardware resources and significantly accelerates performance in large-scale multi-file processing scenarios. In the multi-file observation scenario, we employed the FRB20201124 files from the FAST-FREX dataset, which contains 125 FITS
sample files. This dataset was used to represent a realistic multi-file processing workload, enabling a comparative evaluation of the batch-processing performance between Heimdall and Heimdall++.

In the experimental setup, we set the DM range to 0–1000 cm$^{-3}$. For the original Heimdall, parallelism refers to the number of concurrently executed processes; to ensure the effectiveness of multi-process execution, we enabled the CUDA Multi-Process Service. For Heimdall++, parallelism was adjusted by modifying the number of \texttt{execute threads} (as shown in Fig.~\ref{fig:heimdall++_pipeline}) through command-line parameters.

During testing in our experimental environment, we observed that when the number of concurrent Heimdall processes exceeded two, execution was interrupted due to insufficient GPU memory. Therefore, for Heimdall, we only report performance measurements up to two processes. In contrast, Heimdall++ achieved stable execution with a maximum parallelism of four on the same hardware; thus, the results in the figure are shown up to four threads for Heimdall++.

\begin{figure}[!t]
\centering
\includegraphics[width=1\linewidth]{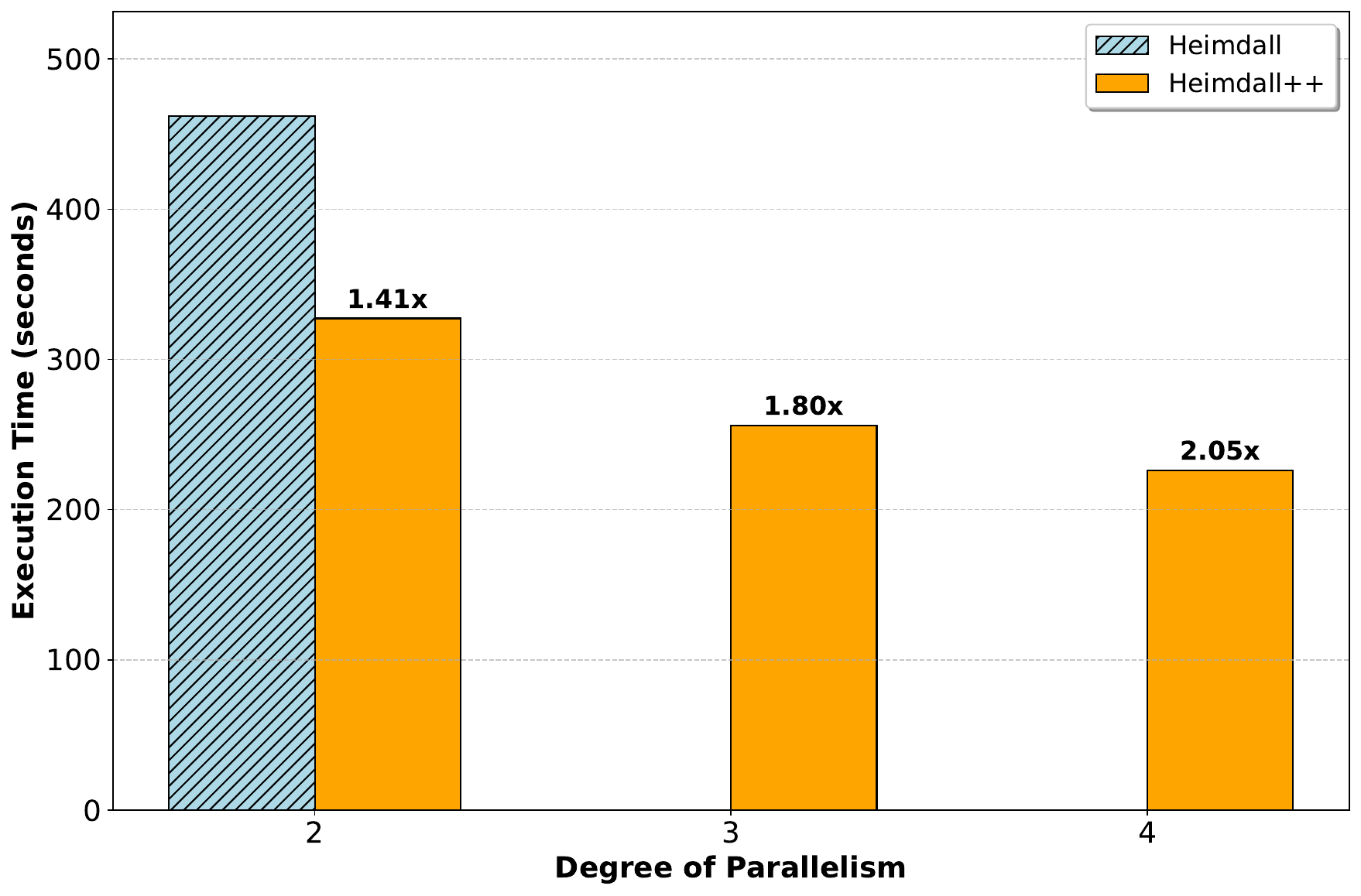}
\caption{Performance comparison between Heimdall and Heimdall++ in various degrees of parallelism}
\label{fig:heimdall_pl}
\end{figure}

As shown in Fig.~\ref{fig:heimdall_pl}, Heimdall++ achieves a 1.41$\times$ speedup over Heimdall at a parallelism level of two, demonstrating the effectiveness of the Heimdall++ pipeline design. By restructuring the division of computation between the CPU and GPU, Heimdall++ enables better concurrency between host and device operations, thereby reducing GPU idle time and improving overall throughput. Furthermore, owing to lower resource contention under multi-threaded concurrency, Heimdall++ supports twice the parallelism achievable with Heimdall. This higher degree of concurrency leads to further performance gains, reaching speedups of 1.80$\times$ and 2.05$\times$ at parallelism levels of three and four, respectively. These results provide clear evidence that Heimdall++ delivers substantial improvements in batch-processing performance for multi-file scenarios. When the degree of parallelism exceeds four, Heimdall++ encounters system resource constraints that may lead to process instability or termination. Consequently, the optimal parallelism level for maximizing throughput is hardware- and workload-dependent. We recommend empirically determining this threshold via benchmarking on representative datasets under target deployment conditions, ensuring optimal resource utilization and system stability.

\section*{Conclusion}
In this study, we have developed and evaluated Heimdall++, an end-to-end optimized redesign of the Heimdall single-pulse search pipeline that addresses key computational bottlenecks limiting GPU utilization. By introducing fine-grained parallelization across CUDA streams, a shared device-memory allocator, and unified memory management, Heimdall++ eliminates redundant host–device data transfers and reduces memory allocation overhead. These optimizations yield up to a 2.66$\times$ speedup over the original Heimdall in single-file processing while preserving full equivalence in search results. For large-scale, multi-file observational scenarios, Heimdall++ further incorporates a multithreaded, pipelined execution framework that decouples CPU-bound pipeline setup from GPU-bound computation. This design mitigates the GPU stall problem and avoids the resource contention inherent in multi-process execution. Evaluated on the FAST-FREX dataset, Heimdall++ achieves up to a 2.05$\times$ acceleration in batch processing and demonstrates superior scalability with increasing concurrency.

These optimizations significantly enhance Heimdall's capability to process large-scale radio astronomy data in real time. This advancement not only reduces computational costs for observatories but also lays the foundation for future high-throughput surveys with next-generation radio telescopes.

\bibliographystyle{IEEEtran}
\bibliography{cites}

@article{staelin1969fast,
  author={Staelin, D.H.},
  journal={Proceedings of the IEEE}, 
  title={Fast folding algorithm for detection of periodic pulse trains}, 
  year={1969},
  volume={57},
  number={4},
  pages={724-725},
  doi={10.1109/PROC.1969.7051}
}

@article{lorimer2007bright,
  title={A bright millisecond radio burst of extragalactic origin},
  author={Lorimer, Duncan R and Bailes, Matthew and McLaughlin, Maura Ann and Narkevic, David J and Crawford, Froney},
  journal={Science},
  volume={318},
  number={5851},
  pages={777--780},
  year={2007},
  publisher={American Association for the Advancement of Science}
}

@article{dewdney2015ska1,
  title={SKA1 system baseline V2 description},
  author={Dewdney, Peter and Turner, W and Braun, R and Santander-Vela, J and Waterson, M and Tan, GH},
  journal={SKA Organisation (November 2015)},
  year={2015}
}

@article{twidle2019impossible,
  title={Impossible images: Radio astronomy, the square kilometre array and the art of seeing},
  author={Twidle, Hedley},
  journal={Journal of Southern African Studies},
  volume={45},
  number={4},
  pages={767--790},
  year={2019},
  publisher={Taylor \& Francis}
}

@article{barsdell2012accelerating,
  title={Accelerating incoherent dedispersion},
  author={Barsdell, Benjamin R and Bailes, Matthew and Barnes, David G and Fluke, Christopher J},
  journal={Monthly Notices of the Royal Astronomical Society},
  volume={422},
  number={1},
  pages={379--392},
  year={2012},
  publisher={The Royal Astronomical Society}
}

@inproceedings{schinckel2012australian,
  title={The australian SKA pathfinder},
  author={Schinckel, Antony E and Bunton, John D and Cornwell, Tim J and Feain, Ilana and Hay, Stuart G},
  booktitle={Ground-based and Airborne Telescopes IV},
  volume={8444},
  pages={807--818},
  year={2012},
  organization={SPIE}
}

@article{amiri2018chime,
  title={The CHIME fast radio burst project: system overview},
  author={Amiri, M and Bandura, K and Berger, P and Bhardwaj, M and Boyce, MM and Boyle, PJ and Brar, C and Burhanpurkar, M and Chawla, P and Chowdhury, J and others},
  journal={The Astrophysical Journal},
  volume={863},
  number={1},
  pages={48},
  year={2018},
  publisher={IOP Publishing}
}

@incollection{price2020real,
  title={Real-time stream processing in radio astronomy},
  author={Price, Danny C},
  booktitle={Big Data in Astronomy},
  pages={83--112},
  year={2020},
  publisher={Elsevier}
}

@article{williamson2024optimising,
  title={Optimising the Processing and Storage of Radio Astronomy Data},
  author={Williamson, Alexander and Elahi, Pascal J and Dodson, Richard and Rhee, Jonghwan and Gong, Qian},
  journal={arXiv preprint arXiv:2410.02285},
  year={2024}
}

@inproceedings{agrawal2024taming,
  title={Taming Throughput-Latency tradeoff in LLM inference with Sarathi-Serve},
  author={Agrawal, Amey and Kedia, Nitin and Panwar, Ashish and Mohan, Jayashree and Kwatra, Nipun and Gulavani, Bhargav and Tumanov, Alexey and Ramjee, Ramachandran},
  booktitle={18th USENIX Symposium on Operating Systems Design and Implementation (OSDI 24)},
  pages={117--134},
  year={2024}
}

@article{ransom2011presto,
  title={PRESTO: pulsar exploration and search toolkit},
  author={Ransom, Scott},
  journal={Astrophysics source code library},
  pages={ascl--1107},
  year={2011}
}

@article{lorimer2011sigproc,
  title={SIGPROC: pulsar signal processing programs},
  author={Lorimer, DR},
  journal={Astrophysics Source Code Library},
  pages={ascl--1107},
  year={2011}
}

@article{men2024transientx,
  title={TransientX: a high-performance single-pulse search package},
  author={Men, Yunpeng and Barr, Ewan},
  journal={Astronomy \& Astrophysics},
  volume={683},
  pages={A183},
  year={2024},
  publisher={EDP Sciences}
}

@article{adamek2020single,
  title={Single-pulse detection algorithms for real-time fast radio burst searches using gpus},
  author={Ad{\'a}mek, Karel and Armour, Wesley},
  journal={The Astrophysical Journal Supplement Series},
  volume={247},
  number={2},
  pages={56},
  year={2020},
  publisher={IOP Publishing}
}

@article{sclocco2016real,
  title={Real-time dedispersion for fast radio transient surveys, using auto tuning on many-core accelerators},
  author={Sclocco, Alessio and van Leeuwen, Joeri and Bal, Henri E and van Nieuwpoort, Rob V},
  journal={Astronomy and computing},
  volume={14},
  pages={1--7},
  year={2016},
  publisher={Elsevier}
}

@article{you2021gpu,
  title={A GPU based single-pulse search pipeline (GSP) with database and its application to the Commensal Radio Astronomy FAST Survey (CRAFTS)},
  author={You, Shan-Ping and Wang, Pei and Yu, Xu-Hong and Xie, Xiao-Yao and Li, Di and Liu, Zhi-Jie and Pan, Zhi-Chen and Yue, You-Ling and Qian, Lei and Zhang, Bin and others},
  journal={Research in Astronomy and Astrophysics},
  volume={21},
  number={12},
  pages={314},
  year={2021},
  publisher={IOP Publishing}
}

@article{mao2025prestozl,
  title={PrestoZL: A GPU-accelerated High-throughput Jerk Search Toolkit for Binary Pulsars},
  author={Mao, Kuang and Tang, Zhaorong and Pan, Qiuhong and Wang, Pei and Chen, Huaxi and Ransom, Scott M and Li, Di and Tang, Xuefei and Wang, Qi and Feng, Yi and others},
  journal={The Astrophysical Journal Supplement Series},
  volume={280},
  number={1},
  pages={36},
  year={2025},
  publisher={IOP Publishing}
}

@article{cordes2003searches,
  title={Searches for fast radio transients},
  author={Cordes, JM and McLaughlin, Maura A},
  journal={The Astrophysical Journal},
  volume={596},
  number={2},
  pages={1142},
  year={2003},
  publisher={IOP Publishing}
}

@article{li2018fast,
  title={FAST in space: considerations for a multibeam, multipurpose survey using China's 500-m aperture spherical radio telescope (FAST)},
  author={Li, Di and Wang, Pei and Qian, Lei and Krco, Marko and Dunning, Alex and Jiang, Peng and Yue, Youling and Jin, Chenjin and Zhu, Yan and Pan, Zhichen and others},
  journal={IEEE Microwave Magazine},
  volume={19},
  number={3},
  pages={112--119},
  year={2018},
  publisher={IEEE}
}

@MISC{freire2024gcpsr,
       author = {{Freire}, P.},
        title = "{Pulsars in globular clusters}",
 howpublished = {\url{https://www3.mpifr-bonn.mpg.de/staff/pfreire/GCpsr.html}},
         year = 2024,
       note = {Accessed: 2025-09-04}
}

@article{Guo_2025,
doi = {10.3847/1538-4365/adf42d},
url = {https://doi.org/10.3847/1538-4365/adf42d},
year = {2025},
month = {sep},
publisher = {The American Astronomical Society},
volume = {280},
number = {1},
pages = {34},
author = {Guo, Xuerong and Wang, Han and Xiao, Yifan and Chen, Huaxi and Ke, Yinan and Miao, ChenChen and Wang, Pei and Li, Di and Jin, Chenwu and He, Ling and Feng, Yi and Zhang, Yongkun and Xu, Jiaying and Chen, Guangyong},
title = {Accelerating the Fast Radio Burst Search: Data Set and Methods},
journal = {The Astrophysical Journal Supplement Series}
}

\end{document}